\documentclass{aastex}
\usepackage{emulateapj5}

\shorttitle{EUV BACKGROUND FROM CLUSTERS OF GALAXIES}
\shortauthors{RANDALL \& SARAZIN}
\slugcomment{Astrophysical Journal, in press}

\begin{document}

\title{The Contribution of EUV from Clusters of Galaxies to the Cosmic Ionizing
Background}

\author{Scott W. Randall and Craig L. Sarazin}

\affil{Department of Astronomy, University of Virginia, 
P. O. Box 3818, Charlottesville, VA 22903-0818;
swr3p@virginia.edu,
cls7i@virginia.edu}

\begin{abstract}
Recent observations with the {\it Extreme Ultraviolet Explorer} ({\it EUVE})
suggest that at least some clusters of galaxies are luminous sources of extreme
ultraviolet (EUV) radiation.
It is not clear yet whether EUV emission is a general feature of clusters;
for the purposes of limiting the contribution to the background radiation,
we assume that it is true of most clusters.
Assuming that the source of the EUV emission is inverse
Compton (IC) scattering of the Cosmic Microwave Background photons by
relativistic electrons, 
we construct a simple model for the expected average emission from clusters as
a function of their mass and the redshift of interest.
Press-Schechter theory is used to determine the abundance of clusters of
various masses as a function of redshift.
We determine the amount of background radiation produced by clusters.
The total mean intensity, spectrum, and the ionization rates for
\ion{H}{1} and \ion{He}{2} are determined at present and at a variety of
redshifts.
Because clusters form by the merger of smaller subclusters, the
amount of EUV background radiation should be larger at present than
in the past.
We compare our results to the ionizing background expected from quasars.
We find that while clusters do contribute a significant EUV background,
it is less than a percent
of that expected from quasars.
\end{abstract}


\section{Introduction} \label{sec:intro}

Recently, extreme ultraviolet (EUV) and very soft X-ray emission
has been detected from a number of clusters of galaxies.
Except for the Coma and Virgo cluster
(Bowyer, Lampton, \& Lieu 1996;
Lieu et al.\ 1996a,b,c;
Bowyer, Bergh\"ofer, \& Korpela 1999;
Bergh\"ofer, Bowyer, \& Korpela 2000),
the individual detections are controversial.
Other clusters which may have been detected include
Abell~1795 and Abell~2199
(Bowyer, Lieu, \& Mittaz 1998;
Mittaz, Lieu, \& Lockman 1998),
Abell~4038
(Bowyer et al.\ 1998),
and Abell~4059
(Bowyer 2000).
The excess EUV detections in the rich clusters (i.e., not Virgo) correspond to
luminosities of $\sim 10^{44}$ ergs s$^{-1}$.
The detections of Abell~1795 and Abell~2199 as extended EUV sources have
been quite controversial.
For both clusters, detections with {\it EUVE} were claimed by
Bowyer et al.\ (1998)
and
Mittaz et al.\ (1998).
However,
Bowyer, Bergh\"ofer, \& Korpela (1999) argued these detections were due to
an incorrect subtraction of the background.
In Abell~2199,
Lieu et al.\ (1999b) reobserved the cluster, and continue to claim
detection of an extended EUV source.
An associated soft X-ray excess may have been seen with {\it BeppoSAX}
(Kaastra et al.\ 1999).
Recently, {\it XMM} appears to have confirmed the presence of very extended,
EUV/soft-X-ray excess emission in Abell~1795
(Arnaud 2000).
In addition to the controversies about individual clusters,
Arabadjis \& Bregman (1999)
argue that all the detections may be due to uncertainties in
Galactic absorption.

Thus, it is not certain at this point whether excess EUV emission is
a general characteristic of clusters of galaxies.
In order to limit the contribution of cluster EUV emission to the
extragalactic background, we will assume that it is, in fact,
a general property of rich clusters.
If this is the case, then our calculations may give a determination of
the cluster contribution to the EUV background.
If EUV excess emission is present in only a fraction of rich clusters,
then our calculations will give a strong upper limit to the contribution
to the extragalactic background.

A promising theory for the EUV emission by clusters is that it is produced
by inverse Compton (IC) scattering of the Cosmic Microwave Background (CMB)
by relativistic electrons in the intracluster medium
(Hwang 1997;
Bowyer \& Bergh\"ofer 1998;
En{\ss}lin \& Biermann 1998;
Sarazin \& Lieu 1998).
This requires the presence of a distinct population of relatively low
energy relativistic electrons with energies of $\sim$100 MeV
(Bowyer \& Bergh\"ofer 1998)
beyond the population of higher energy electrons which produce
radio emission in a small fraction of clusters
(e.g., Feretti 1999).
Such a distinct population is expected theoretically, because the
lower energy electrons have lifetime which are comparable to the
Hubble time, while the radio emitting electrons have very short
lifetimes
(Sarazin 1999;
Atoyan \& V\"olk 2000;
Takizawa \& Naito 2000).

An alternative theory is that the EUV emission is thermal in origin
(Mittaz et al.\ 1998;
Lieu, Bonamente, \& Mittaz 1999a).
There are a number of concerns with this thermal model, particularly
with the energetics
(Fabian 1996;
Sarazin \& Lieu 1998).
Recently, Lieu et al.\ (1999a) have argued that the observed EUV to soft
X-ray spectra of clusters decline rapidly with increasing frequency,
and that this favors a thermal model.
However, such a rapid decline is, in fact, what was predicted
by the IC model
(Sarazin \& Lieu 1998;
Sarazin 1999;
Atoyan \& V\"olk 2000;
Takizawa \& Naito 2000;
Figure~\ref{fig:spectrum_model} below).

In any case, we will normalize our calculations of the EUV emission
from clusters to the observed EUV flux from the Coma cluster, which is
the best observed rich cluster.
Thus, our results are somewhat independent of the emission mechanism
for the EUV emission.
In detail, we need a model to scale the EUV luminosity of Coma to other
clusters and other redshifts, and we need a model for the spectrum
of the emission.
At present, there are no detailed thermal models for the EUV emission
from clusters which provide such predictions.
On the other hand, the IC model provides a very simple scaling
of the luminosity and spectrum to other masses of clusters and to other
redshifts
(\S~\ref{sec:emit} below).
Thus, we will adopt the IC model for the emission to scale the observed
flux of the Coma cluster to other clusters and redshifts.

If this radiation really is a common feature of clusters at the indicated
luminosity levels, it might make a contribution to the intergalactic
background radiation field at UV through soft X-ray energies.
In this paper, we estimate the mean intensity of the 
diffuse EUV background from galaxy clusters, both at the
current epoch and at higher redshifts.
We will compare the radiation field from clusters with that
produced by quasars and other active galactic nuclei (AGN)
(e.g., Haardt \& Madau 1996).
In \S~\ref{sec:mass}, we use Press-Schechter (1974) theory to calculate
the abundance of clusters of a given mass as a function of redshift.
In \S~\ref{sec:emit}, the EUV emission of clusters of a given mass is
calculated, based on a simple IC theory.
We assume that both the thermal energy of the intracluster gas and the 
energy of relativistic electrons are produced ultimately by intracluster
shocks, and that these shocks convert a fixed fraction of the shock energy
into relativistic particles.
The resulting diffuse radiation field at the present epoch and at
higher redshifts are given in \S~\ref{sec:results}.
The resulting ionization rates of \ion{H}{1} and \ion{He}{2} are
also derived there.
The cluster diffuse EUV is compared to that produced by quasars and
other AGN.
The conclusions are summarized in \S~\ref{sec:conclusion}.

\section{Mass Distribution of Clusters} \label{sec:mass}

\subsection{Predicted Mass Distribution of Clusters} \label{sec:model}

In order to determine the total background radiation from galaxy
clusters, it is first necessary to model their number density and its
evolution with redshift $z$.
While the present-day abundance of rich clusters is well-determined from
observations
(see \S~\ref{sec:mass_compare} below),
the evolution of the cluster abundance is poorly determined
at present, particularly at high redshifts.
Thus, we will use the Press-Schechter (PS) formalism
(Press \& Schechter 1974)
to model the cluster density and its evolution, with parameters chosen
to fit the observed present-day properties of clusters and the results
of more detailed numerical models of large-scale structure.
Comparisons to observations of clusters and to numerical simulations
show that the PS provides an excellent representation of the statistical
properties of clusters, if the PS parameters are carefully selected
(e.g., Lacey \& Cole 1993;
Bryan \& Norman 1998;
Viana \& Liddle 1999).
Let $n(M,z) dM$ be the comoving number density of clusters with masses in
the range $M$ to $M + d M$ in the Universe at a redshift of $z$.
According to PS, this differential number density is given by
\begin{eqnarray} \label{eq:psfunc}
n(M,z) \, dM & = & \sqrt{ \frac{2}{\pi}} \, \frac{ \overline{\rho}}{M^{2}}
\, \frac{\delta_{c}(z)}{\sigma (M) } \,
\left| \frac{d \, \ln \, \sigma (M) }{d \, \ln \, M} \right|
\nonumber \\
& & \times \, \exp \left[- \frac{\delta_{c}^{2}(z)}{2\sigma^{2} (M) } \right]
\, dM
\, ,
\end{eqnarray}
where $\overline{\rho}$ is the current mean density of the Universe,
$\sigma(M)$ is the current rms density fluctuation within a sphere of
mean mass $M$,
and $\delta_{c}(t)$ is the critical linear overdensity for a region to
collapse.
In using this expression, we have made the usual assumptions 
that each cluster under consideration has recently collapsed
$(z_{obs} \approx z_{coll})$ and that collapse is spherically
symmetric.
However, comparisons to numerical simulations show that this expression
gives a good fit to the detailed results if the parameters are selected
carefully
(e.g., Lacey \& Cole 1993;
Viana \& Liddle 1999).
We will assume a power-law spectrum of density perturbations,
which is consistent with CDM models:
\begin{equation}  \label{eq:sigma}
\sigma (M) = \sigma_{8} \, \left( \frac{M}{M_{8}} \right) ^{-\alpha} \, ,
\end{equation}
where $\sigma_{8}$ is the present day rms density fluctuation on a scale of
8 $h^{-1}$ Mpc,
$M_8 = ( 4 \pi / 3 ) ( 8 \, h^{-1} \, {\rm Mpc} )^3 \bar{\rho}$
is the mass contained in a sphere of radius 8 $h^{-1}$ Mpc, 
and the Hubble constant is
$H_0 = 100 \, h$ km s$^{-1}$ Mpc$^{-1}$.
The exponent $\alpha$ is given by $\alpha = \frac{n+3}{6}$, where the power
spectrum of fluctuations varies with wavenumber $k$ as $k^n$;
we assume $n=-7/5$, which leads to $\alpha = 4/15$ throughout.
Using the above relations, we can rewrite
equation~(\ref{eq:psfunc}) as
\begin{eqnarray} \label{eq:psfunc2}
n(M,z) \, dM & = & \sqrt{ \frac{2}{\pi}}\, \frac{ \overline{\rho}}{M^{2}}
\, \frac{\delta_{c}(z)}{\sigma (M) } \,
\left| \alpha \right|
\nonumber \\
& & \times \, \exp \left[-
\frac{\delta_{c}(z)^{2}}{2\sigma^{2}} \right]
\, dM
\, .
\end{eqnarray}

The normalization of the power spectrum and overall abundance of clusters
is set by $\sigma_8$.
Although there are published values for this constant
(e.g.,
Viana \& Liddle 1999;
Bahcall \& Fan 1998),
we choose to normalize our calculations to a fixed value for the observed
local abundance of clusters.
Specifically, for each cosmology,
we choose the value of $\sigma_{8}$ such that
the present day integrated number density of clusters
with mass
$M > 8 \times 10^{14} \, h^{-1} \, M_{\odot}$
matches the observed value of
$2 \times 10^{-7} \, h^{3}$ Mpc$^{-3}$
(Bahcall \& Fan 1998).
This method has the advantage that it forces the
present day cluster abundance to be the same for each cosmology, as
opposed to having a small spread in values, which occurs if an analytic
approximation for the variation of $\sigma_{8}$ with cosmology is used.

The cosmological model is characterized by several parameters.
The dependence on the Hubble constant
$H_0 = 100 \, h$ km s$^{-1}$ Mpc$^{-1}$
just produces an overall scaling of the results, which we include through
the factor $h$.
We characterize the cosmological solution by the two standard dimensionless
parameters.
First, $\Omega_0 \equiv \bar{\rho} / {\rho_c} $ is the ratio of the current mass
density to the critical mass density $\rho_c = 3 H_0^2 / ( 8 \pi G )$.
Second, $\Omega_{\Lambda} \equiv \Lambda / ( 3 H_0^2 )$, where 
$\Lambda$ is the cosmological constant.
We will consider three distinct cosmological models in our analysis:
an Einstein-deSitter universe ($\Omega_0=1$, $\Omega_{\Lambda}=0$),
an open universe ($\Omega_0=0.3$, $\Omega_{\Lambda}=0$),
and a low-density flat universe
($\Omega_0=0.3$, $\Omega_{\Lambda}=0.7$).
The number density of clusters evolves differently in each of these three
models.
While the latter two models agree better with observations
(e.g., Bahcall \& Fan 1998),
we include the closed model for completeness.

The evolution of the density of clusters is
encapsulated in the $\delta_{c}(z)$ term in
equation~(\ref{eq:psfunc2}).
In general, $\delta_c (z) \propto D(t)$, where
$D(t)$ is the growth factor of linear perturbations as a function of
cosmic time  $t$
(see Peebles [1980], \S~11 for details).
Expressions for the $\delta_c (z)$ in different cosmological models
are:
\begin{equation}   \label{eq:delta}
\delta_{c}(z) = \left\{
\begin{array}{ll} 
\frac{3}{2} D( t_0 ) \left[ 1 + 
\left( \frac{t_{\Omega}}{t}
\right)^{\frac{2}{3}}
\right]    \\
\frac{3(12\pi)^\frac{2}{3}}{20} \left(
\frac{t_0 }{t}
\right)^{\frac{2}{3}}  \\
\frac{D(t_0 )}{D(t)} 
\left(\frac{3(12\pi)^\frac{2}{3}}{20}\right) 
\left(1 + 0.0123 \ \log_{10}\Omega_{f}
\right)
\end{array}
\right.
\end{equation}
for the open, closed, and flat models respectively.  For the open model ($\Omega_0 < 1$, $\Omega_\Lambda = 0$),
$t_{\Omega} \equiv
\pi H_0^{-1} \Omega_0 \left( 1 - \Omega_0 \right)^{-\frac{3}{2}}$
represents the epoch at which a nearly constant expansion takes over
and no new clustering can occur, and the growth factor can be expressed as
\begin{equation}  \label{eq:growth_factor}
D(t)=\frac{3\ \sinh \eta \left( \sinh \eta - \eta \right)}
	  {\left( \cosh \eta - 1 \right)^{2}} - 2
\end{equation} 
where $\eta$ is the standard parameter in the cosmic expansion equations
(Peebles 1980, eqn.~13.10)
\begin{equation} \label{eq:eta}
\begin{array}{ll}
\frac{1}{1+z} = \frac{\Omega_0}
{2 \left( 1 - \Omega_0 \right)}
\left( \cosh \eta - 1 \right) \, , \\
H_0 t = \frac{\Omega_0 }
{2 \left( 1 - \Omega_0 \right)^{\frac{3}{2}}}
\left( \sinh \eta - \eta \right) \, .
\end{array}
\end{equation}
The solution for $\delta_{c}$ in the Einstein-deSitter model
can be obtained from the open model solution by the limit
$t_{\Omega}/t \rightarrow \infty$
(Lacey \& Cole 1993).
To evaluate $\delta_{c}$ in the flat model
$(\Omega_0 + \Omega_{\Lambda} = 1$),
we have used an approximation given by Kitayama \& Suto (1996).
Here $\Omega_{f}$ is the value of the mass density ratio $\Omega$ at
the redshift of formation,
\begin{equation} \label{eq:omega_f}
\Omega_{f} = \frac{\Omega_0
	\left( 1+z \right)^{3}}
	{\Omega_0 \left( 1+z \right)^{3} + \Omega_{\Lambda}} \, .
\end{equation}
In this model the growth factor can be written as
\begin{equation} \label{eq:growth_flat}
D(x)= \frac{(x^{3} + 2)^{1/2}}{x^{3/2}} \,
\int_{x_0}^{x} x^{3/2} \, (x^{3}+2)^{-3/2}dx
\end{equation}
(Peebles 1980, eqn.~13.6) where $x_0 \equiv ( \frac{2
\Omega_{\Lambda}}{\Omega_0} )^{1/3} $ and $x = x_0/(1+z)$.

\begin{figure*}[p]
\vskip5.3truein
\includegraphics{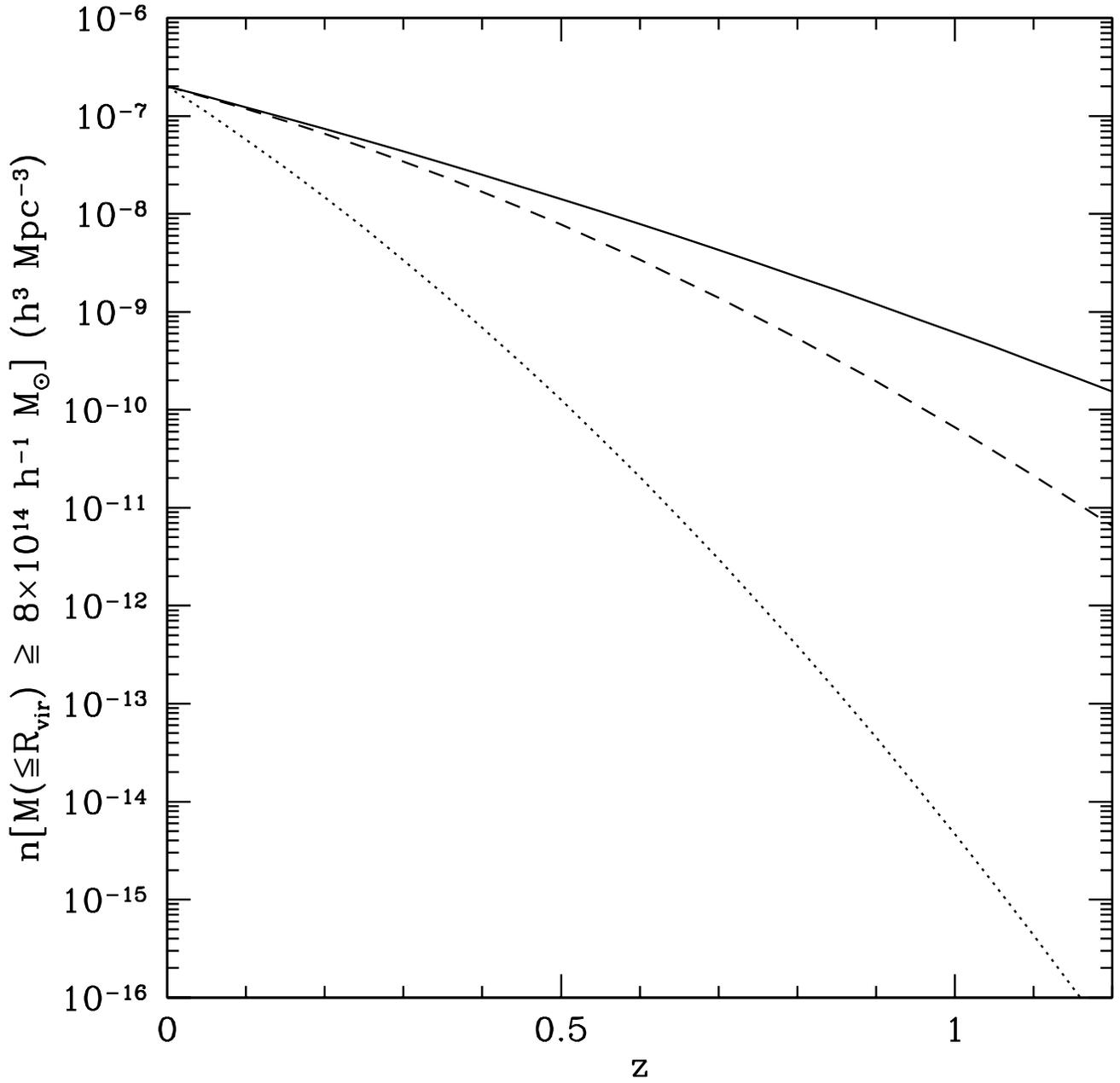}
\caption[fig1.ps]{
The evolution of the abundance of clusters with redshift $z$ for three
distinct cosmologies.
Only clusters with mass $M \geq 8 \times 10^{14} \, h^{-1} \, M_{\odot}$
within their virial radius $R_{vir}$ are considered.  The
solid curve represents an open Universe ($\Omega_0 = 0.3 \, , \,
\Omega_{\Lambda}=0 $), the dashed curve a flat Universe ($\Omega_0 =
0.3 \, , \,
\Omega_{\Lambda} = 0.7$), and the dotted curve a closed Universe
($\Omega_0 = 1.0 \, , \,
\Omega_{\Lambda} = 0$).
\label{fig:mass_dist}}
\end{figure*}

Figure~\ref{fig:mass_dist} shows how the cluster abundance evolves
with redshift for the three cosmologies under consideration.
For each model we plot the integrated number density of clusters at each
redshift, which is simply the integral of equation~(\ref{eq:psfunc2})
over all masses 
$M \geq 8 \times 10^{14} \, h^{-1} \, M_{\odot}$.
The mass plotted here is the virial mass, which is the mass within the
virial radius.
The closed model shows the strongest evolution, followed by the flat
model, with the open model showing the weakest evolution, as expected
(e.g., Bahcall \& Fan 1998).

\subsection{Comparison to Observed X-ray Luminosity Function}
\label{sec:mass_compare}

\begin{figure*}[p]
\vskip5.3truein
\includegraphics{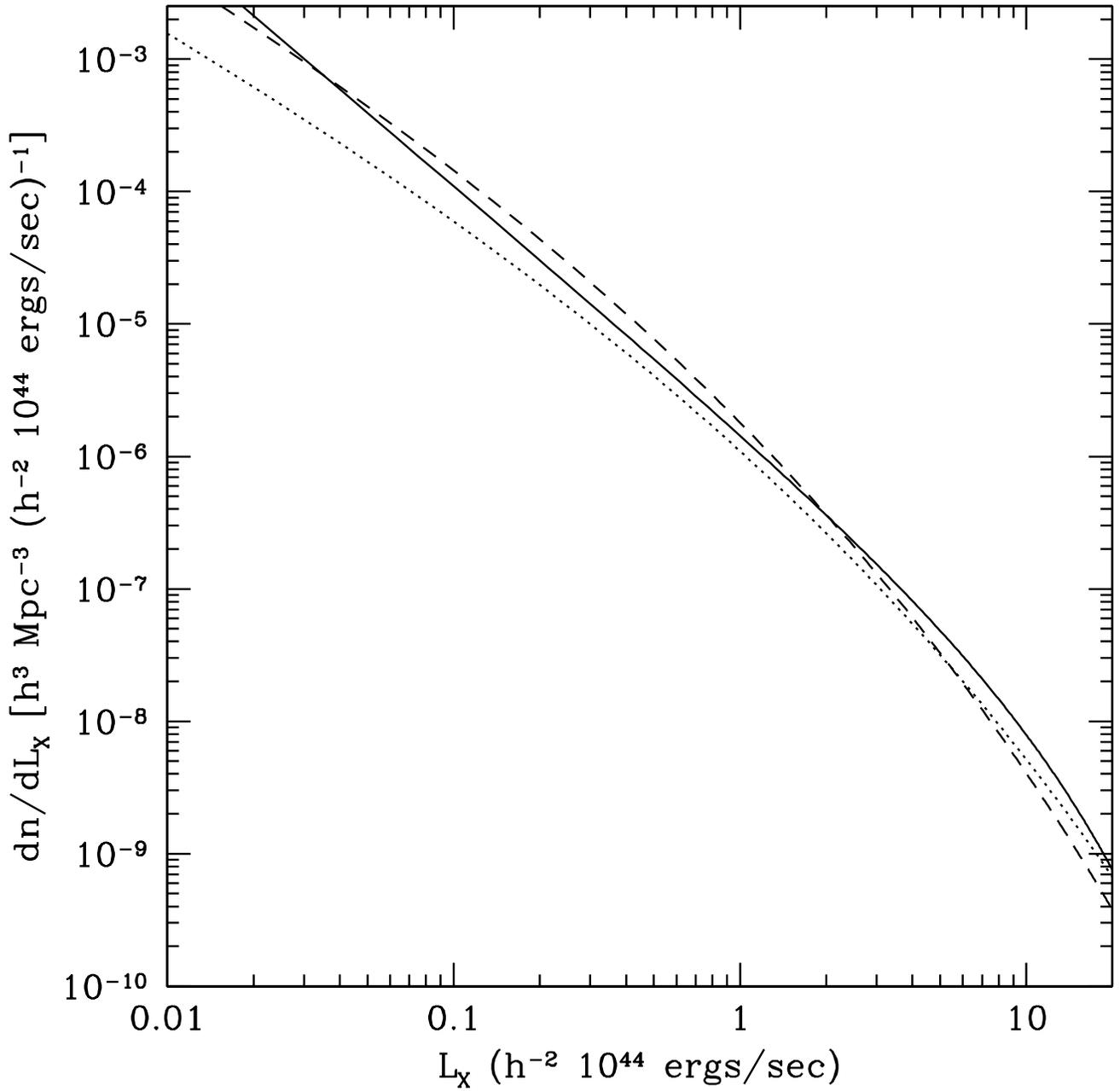}
\caption{
The present day cluster bolometric X-ray luminosity function.
The solid curve is the observed luminosity function
(Ebeling et al.\ 1997).
The dashed curve shows the predicted luminosity function for a closed
Universe
($\Omega_0 = 1$, $\Omega_{\Lambda}=0$),
while the dotted curve shows the
predicted luminosity function for an open Universe
($\Omega_0 = 0.3$, $\Omega_{\Lambda} = 0$).
The flat Universe is omitted here 
for clarity since the predicted curve essentially matches the
predicted curve for the open Universe.
\label{fig:diff_lum}}
\end{figure*}

As a check on these models, we have estimated the resulting X-ray luminosity
function of clusters at the present epoch.
To convert the predicted mass function of clusters to an X-ray luminosity
function, we first determine the temperatures of the gas in clusters
using the observed mass-temperature relationship
(Evrard, Metzler, \& Navarro 1996;
Horner, Mushotzky, \& Scharf 1999)
\begin{equation} \label{eq:mass_temp}
M = c_0 \times 10^{13} \, h^{-1} \, \left( \frac{T_X}{T_m} \right)^{c_1}
\, M_\odot \, ,
\end{equation}
where $T_m \equiv 1$ keV, $c_0 \approx 5$, and $c_1 \approx 1.5$.
Then, the
X-ray luminosities $L_X$ were calculated from the X-ray temperatures using
the the X-ray luminosity-temperature relationship
(Arnaud \& Evrard 1999)
\begin{equation} \label{eq:xray_lum_temp}
L_X = b_0 \times 10^{44} \, h^{-2} \, \left( \frac{T_X}{T_l} \right)^{b_1}
\, {\rm erg} \, {\rm s}^{-1} \, ,
\end{equation}
where $T_X$ is the X-ray temperature, $T_l \equiv 6$ keV is a characteristic
value used for scaling, $d_0 \approx 2.88$, and $d_1 \approx 2.88$.
Here, $L_X$ is the bolometric X-ray luminosity.
Figure~\ref{fig:diff_lum} shows the resulting differential luminosity
function at the current epoch as predicted by PS for our three cosmologies.

For comparison, we overplot the observed local bolometric X-ray luminosity
function from
Ebeling et al.\ (1997),
\begin{eqnarray} \label{eq:xraylumfunc}
\frac{d n}{d L_X} & = & a_0 \times 10^{-8} \, h^5 \,
\exp ( - L_X / L^*_X ) 
\nonumber \\
& & \times \, \left( \frac{L_X}{L^*_X} \right)^{- a_1}
\, {\rm Mpc}^{-3} \, ( 10^{44} \, {\rm erg} \, {\rm s}^{-1} ) \, .
\end{eqnarray}
Here, $L^*_X \approx 9.3 \times 10^{44}$ erg s$^{-1}$ is a characteristic
cluster X-ray luminosity,
$a_0 \approx 2.64$, and $a_1 \approx 1.84$.
The model is in reasonable agreement with the observations.

\section{Predicted Emission from Clusters of Galaxies} \label{sec:emit}

As discussed in \S~\ref{sec:intro}, EUV and very soft X-ray emission
has been detected from a number of clusters of galaxies.
Although it is not yet certain how common this emission is, we
will assume that it is a general feature of clusters.
As we noted in \S~\ref{sec:intro}, this implies that our results are
upper limits on the extragalactic EUV background due to clusters, if
EUV emission is not a common phenomena.
While the mechanism for this emission is still under debate,
inverse Compton (IC) scattering of the cosmic microwave
background (CMB) by cosmic-ray electrons
seems to be a very strong candidate
(Hwang 1997;
Bowyer \& Bergh\"ofer 1998;
En{\ss}lin \& Biermann 1998;
Sarazin \& Lieu 1998).
For the purposes of this analysis, we will assume that IC
scattering is the dominant source of EUV radiation from clusters.
However, we will normalize our models to the observed EUV flux of
the Coma cluster;
the IC model will be used to scale this flux to other cluster masses
and redshifts.

The IC luminosity from an individual cluster is given by
\begin{equation}  \label{eq:lic}
L_{IC} = \frac{4}{3} \, \frac{\sigma_{T}}{m_{e}c}
\langle \gamma \rangle U_{CMB}E_{CR},
\end{equation}
where $U_{CMB}$ is the energy density in the CMB, $E_{CR}$ is the
total energy in cosmic-ray electrons in the cluster, and
$\langle \gamma \rangle$ is the average Lorentz factor of the cosmic-rays:
\begin{equation}  \label{eq:avg_gamma}
\langle \gamma \rangle \equiv \frac
{\int \, n_{CR}(\gamma)\gamma^{2}d\gamma}
{\int \, n_{CR}(\gamma) \gamma d\gamma},
\end{equation}
where $n_{CR}(\gamma)d\gamma$ is the number density of cosmic-ray
electrons with Lorentz factors between $\gamma$ and $\gamma\ +\
d\gamma$.
The {\it EUVE} observations of clusters suggest that
$\langle \gamma \rangle \approx 300$
(for a Hubble constant of 50 km s$^{-1}$ Mpc$^{-1}$;
Sarazin \& Lieu 1998).
One expects that this value will correspond to electrons whose
lifetimes are similar to the age of the cluster
(Sarazin 1999), which implies that $\langle \gamma \rangle \propto h$.
Thus, we assume that
\begin{equation}  \label{eq:avg_gamma_2}
\langle \gamma \rangle  = 600 \, h 
\end{equation}
(Sarazin \& Lieu 1998).
The energy density in the CMB is well determined by observations,
$U_{CMB} = 4 \, \sigma_{T} T_{CMB}^{4} / c
= 4.20 \times 10^{-13} (1 + z)^4$
ergs cm$^{-3}$, where $T_{CMB}$ is the temperature of the CMB, 
$T_{CMB} = 2.73 (1 + z) $ K.  

The observed EUV luminosities of clusters imply that the total
energies of cosmic ray electrons are $E_{CR} \sim 10^{62}$ ergs
(for a Hubble constant of 50 km s$^{-1}$ Mpc$^{-1}$;
Sarazin \& Lieu 1998).
These values are typically 1--10\% of the thermal energy content of
the intracluster gas, $E_{gas}$. 
The observed values of $E_{gas}$ scale as $E_{gas} \propto h^{-5/2}$.
The thermal energy content of the intracluster gas is the result of shock
heating, where the shocks might include cluster merger shocks, overall
accretion shocks, and shocks produced by galaxy winds
(Sarazin 2000).
These shocks all have velocities of $\sim$1000 km s$^{-1}$.
Supernova remnants and other diffuse astrophysical shocks with similar
velocities accelerate relativistic particles,
and the radio properties
of supernova remnants imply that $\sim$3\% of the shock energy goes
into accelerating relativistic electrons
(e.g., Blandford \& Eichler 1987).
Since both the relativistic electrons and the thermal energy in the
intracluster gas arise from the same shocks, 
it seems reasonable to assume that the energy which is injected in
relativistic electrons is proportional to the thermal energy content
of the ICM.
Thus, we will assume that
\begin{equation}  \label{eq:ECR}
E_{CR} = f_{CR} E_{gas} \, .
\end{equation}
We adopt a fixed value for $f_{CR}$ which is determined by the observed
EUV flux from the Coma cluster, which is the best observed rich cluster.
To determine $f_{CR}$, we apply the Galactic absorption and redshift
appropriate to Coma to our adopted spectral model
(see Figure~\ref{fig:spectrum_model} below).
Then, the resulting spectrum was convolved with the {\it EUVE} effective
area to give the expected count rate.
This was compared to the observed values of the {\it EUVE} count rate
(Lieu et al. 1996a;
Bowyer \& Bergh\"ofer 1998;
Bowyer et al. 1999) which range over a factor of about two.
The model spectrum was renormalized to give the observed count rate,
and the total cosmic ray energy $E_{CR}$ was determined.
The {\it ROSAT} X-ray observations of Coma were used to determine the
total gas mass within the projected radius of the {\it EUV} observations
(e.g., Mohr, Mathiesen, \& Evrard 1999).
The corresponding thermal energy content of the gas $E_{gas}$ was derived.
Finally, $E_{CR}$ and $E_{gas}$ for Coma were compared to give $f_{CR}$
(eq.~\ref{eq:ECR}).
The values were $f_{CR} = ( 0.014 - 0.031 ) h^{-1/2}$, corresponding to 
the range in observed count rates for Coma.
We adopt the value $f_{CR} = 0.02 \, h^{-1/2}$.
This is consistent with the reported fluxes of other clusters
(Sarazin \& Lieu 1998).
For any individual cluster, it is likely that $E_{CR}$ and $f_{CR}$ will
depend on its dynamical history, but we assume that on average there is
a simple relationship between the thermal energy in the gas and the
energy in relativistic particles.

The ICM thermal energy is given by
\begin{equation}  \label{eq:Egas}
E_{gas} = \frac{3}{2} \, \frac{k T}{\mu m_p} \, M_{gas} \, ,
\end{equation}
where $T$ is the average temperature of the ICM, $\mu = 0.61$ is the mean
mass per particle in units of the proton mass $m_p$, and $M_{gas}$ is
the mass of the ICM.
There are a number of arguments which suggest that a portion of the
thermal energy of the ICM may have come from non-gravitational effects
which probably occurred before the clusters formed
(e.g., Kaiser 1991;
Cavaliere, Menci, \& Tozzi 1997).
Even if this heating involved shocks, it is unlikely that primary electrons
accelerated in such shocks survive to the present epoch
(Sarazin 1999).
Thus, we will determine the gas temperature in our model clusters using
a theoretical relationship based on purely gravitational effects,
which is consistent with the PS formulation and with numerical hydrodynamical
simulations of cluster formation from large scale structure.
We choose the mass-temperature relationship
\begin{eqnarray} \label{eq:mass_temp_model}
kT & = & 1.39 \, f_{T} \,
\left( \frac{M}{10^{15} \, M_{\odot}} \right)^{2/3}
\nonumber \\
& & \times \, \left[ h^2 \, \Delta_{c} \, E(z)^2 \right]^{1/3} \, {\rm keV} \, ,
\end{eqnarray}
where $f_{T} \approx 0.8$ is a normalization factor
(Bryan \& Norman 1998).
The quantity $E(z)$ is defined as $E(z)^2 \equiv \Omega_0 (1+z)^3 + 
\Omega_{R}(1+z)^2+\Omega_{\Lambda}$, where $\Omega_{R} \equiv 1/(H_0 R)^2$
where $R$ is the current radius of curvature of space.
$\Delta_{c}$ represents the mean density of the cluster divided by the
critical density at a given redshift.
Fits for this parameter are given by Bryan \& Norman
(1998) for the relevant cosmological models:
\begin{equation} \label{eq:Delta_c}
\Delta_{c} = \left\{ \begin{array}{ll}
			18\pi^2\,+\,82x\,-\,39x^2  & 
				(\Omega_{R} = 0) \\
			18\pi^2\,+\,60x\,-\,32x^2  &
				(\Omega_{\Lambda}=0)
			\end{array} \right.
\end{equation}
where $x \equiv [\Omega_0 (1+z)^3/E(z)^2] - 1$.  
Note that
$\Omega_{R} = 0$ is a consequence of the flat model (flat space has an 
infinite curvature).  When $\Omega_0 = 1$, $ x=0$, giving
$\Delta_{c} = 18\pi^2$.

Finally, we need to determine the total gas mass $M_{gas}$ of our
clusters.
X-ray observations indicate that the gas fraction in cluster is
fairly constant for rich clusters, with a value of $\approx$22\% 
(for a Hubble constant of 50 km s$^{-1}$ Mpc$^{-1}$;
e.g., Arnaud \& Evrard 1999).
Thus, we assume that
\begin{equation}  \label{eq:Mgas}
M_{gas} = f_{gas} \, M \, ,
\end{equation}
with $f_{gas} = 0.07 \, h^{-3/2}$.

Using equations~(\ref{eq:lic})-(\ref{eq:Mgas}),
we calculate the average IC luminosity $L_{IC}$ of a cluster of a given total
mass $M$.
For individual clusters, one expects the spectrum of IC EUV emission will
depend on the history of particle acceleration in the cluster
(Sarazin 1999).
For this study, we adopt a ``typical'' IC EUV spectrum, since we intend
to average over an ensemble of clusters to determine the cluster
contribution to the EUV	background.
Specifically, we assume the spectral shape given for Model~11 in
Sarazin (1999).
This adopted model spectrum $L_\nu$(model) is shown in
Figure~\ref{fig:spectrum_model}, where
$L_\nu$ is the luminosity per unit frequency $\nu$.
This particular model spectrum has a total IC luminosity of
$L_{IC} ({\rm model}) = 2.86 \times 10^{44}$ ergs s$^{-1}$.
The normalization of the spectrum of each model cluster is scaled to give
the correct value of $L_{IC}$ for that cluster.
At redshifts $z > 0$, the spectrum is blue shifted due to the increase in
the temperature of the CMB.
Thus, the spectrum of emission of a cluster with total mass $M$
at redshift $z$ is given by
\begin{equation}  \label{eq:spectrum}
L_{IC} ( M , z , \nu ) =  L_{IC} ( M, z) \,
\frac{ L_{\nu / ( 1 + z )} ({\rm model})}{(1 + z ) L_{IC} ({\rm model})}
\, .
\end{equation}

\subsection{The Diffuse Radiation Field} \label{sec:rad_field}

The total emissivity of cluster IC emission at any given redshift is
found by integrating the emission due to clusters of a given total
mass $M$ (eq.~\ref{eq:spectrum})
over the PS mass function (eq.~\ref{eq:psfunc2}):
\begin{equation}  \label{eq:emissiv2}
\epsilon (\nu, z)  = \int \, L_{IC}( M, z, \nu) \, n(M, z) \, dM \, .
\end{equation}
The total emissivity over all frequencies is just
\begin{equation}  \label{eq:emissiv}
\epsilon_{tot} (z) = \int \, \epsilon (\nu, z) \, d \nu 
= \int \, L_{IC} (M, z) \, n(M, z) \, dM \, .
\end{equation}

The equation of radiative transfer for the emission gives a mean intensity
for the radiation field at an observed redshift $z_o$ and observed
frequency $\nu_o$ of
(Haardt \& Madau 1996)
\begin{eqnarray} \label{eq:mean_intensity}
J(\nu_o, z_o) & = & \frac{1}{4 \pi} \int_{z_o}^{\infty}
			dz \, \frac{dl}{dz} \frac{(1+z_o)^3}{(1+z)^3}
\nonumber \\			
& & \times \, \epsilon(\nu,\,z) \,
			e^{-\tau_{eff}(\nu_o, z_o, z)} \, .
\end{eqnarray}
Here, $dl/dz$ is the line element in a Friedmann cosmology, given
by
\begin{eqnarray} \label{eq:dl_dz}
\frac{dl}{dz} & = & \frac{c}{H_0} \, (1+z)^{-1}
\nonumber \\
& & \times \left[ \Omega_0 (1+z)^{3}+\Omega_{R}(1+z)^{2}+\Omega_{\Lambda}
\right]^{-1/2} \, .
\end{eqnarray}
The quantity $\tau_{eff} ( \nu_o , z_n , z )$ gives the effective optical
depth of the intergalactic medium (IGM) at the observed frequency between
$z_o$ and $z$.
We can simplify
equation~(\ref{eq:mean_intensity}) by noting that the energy density
in the CMB, which is present in our expression for $\epsilon(z)$ through
the $L_{IC}$ term, goes as $(1+z)^4/(1+z_o)^4$.  When we consider the
emissivity per unit frequency $\epsilon(\nu, z)$, we get another
factor of $(1+z_o)/(1+z)$ due to the redshift applied at each
frequency.
Multiplying these factors together gives us a net
factor of $(1+z)^3/(1+z_o)^3$, which directly cancels the redshift term in
equation~(\ref{eq:mean_intensity}).
We can therefore rewrite this equation as
\begin{equation} \label{eq:mean_intensity2}
J(\nu_o , z_o) = \frac{1}{4\pi} \int_{z_o}^{\infty}
		dz \, \frac{dl}{dz} \, \epsilon ( \nu_o , z )
		\, e^{-\tau_{eff}(\nu_o, z_o, z)} \, ,
\end{equation}
where $\epsilon(\nu_o, z)$ is defined as the emissivity for a cluster 
population at a redshift $z$ calculated as though it were actually
located at redshift $z_o$.

The opacity of the IGM is mainly due to the effective column density of the less
ionized clouds (such as Ly$\alpha$ clouds).
In determining the optical depth of the IGM, we closely follow the treatment 
in Haardt \& Madau (1996).
They write the optical depth of the IGM as
\begin{eqnarray}
\tau_{eff} (\nu_o, z_o, z) & = & \int_{z_o}^{z}  dz' \int_0^{\infty}
dN_{HI} \, \frac{\partial^{2}N}{\partial N_{HI} \partial z'} \,
\nonumber \\
& & \times \left[ 1 - e^{-\tau ( N_{HI} , z', \nu )} \right] \, ,
\end{eqnarray}
where $N_{HI}$ is the column density of neutral hydrogen in a IGM cloud,
and 
$( \partial^{2}N / \partial N_{HI} \partial z' )$ gives the number of
clouds per unit \ion{H}{1} column and per unit redshift $z'$.
The quantity
$\tau ( N_{HI} , z', \nu )$ is the optical depth of a cloud with
an \ion{H}{1} column of $N_{HI}$ at redshift $z'$ for the frequency
$\nu = \nu_o ( 1 + z' ) / ( 1 + z_o )$;
this depends on the helium abundance and the ionization structure
of the clouds;
we adopt the abundances and ionization structures calculated in
Haardt \& Madau (1996).
For the ionization, we use the approximation given in their
equation~(12),  with $\eta_{thin}$ taken from their Figure 7.

\section{Results} \label{sec:results}

\subsection{Models without IGM Absorption} \label{sec:noabs}

We first calculate the diffuse radiation field neglecting IGM absorption
($\tau_{eff} = 0$ in eq.~[\ref{eq:mean_intensity}]).
Here, we consider only the values at the present epoch, $z_o = 0$.
We find the total mean intensity to be
$1.0 \times 10^{-10}$ ergs cm$^{-2}$ s$^{-1}$ sr$^{-1}$
for the open model,
$9.0 \times 10^{-11}$ ergs cm$^{-2}$ s$^{-1}$ sr$^{-1}$
for the flat model, and
$ 4.7 \times 10^{-11}$ ergs cm$^{-2}$ s$^{-1}$ sr$^{-1}$
for the closed model.

We have calculated the spectrum of the cluster EUV background at the 
present epoch, $z=0$, neglecting absorption for our three
cosmological models.  Results are plotted in Figure~\ref{fig:spec_lum_noabs}.
The corresponding ionization rates have been determined by integrating 
each of these radiation fields over the ionization cross-sections of
\ion{H}{1} and \ion{He}{2}.  The results are
$\zeta_H = 1.8 \times 10^{-17}$,
$1.6 \times 10^{-17}$,
$8.6 \times 10^{-18}$ and
$\zeta_{HeII} = 2.05 \times 10^{-18}$,
$1.8 \times 10^{-18}$,
$9.7 \times 10^{-19}$ s$^{-1}$
for the open, flat, and closed cosmological models, respectively.

\subsection{Models with IGM Absorption} \label{sec:abs}

Next, we calculate models including the opacity of the IGM, following the
treatment of Haardt \& Madau (1996).

\subsubsection{Present Epoch Diffuse Radiation}

We have first calculated the values for the diffuse cluster IC radiation
field at the present epoch, $z_o = 0$. 
We find the total mean intensity to be
$9.0 \times 10^{-11}$ ergs cm$^{-2}$ s$^{-1}$ sr$^{-1}$
for the open model,
$8.2 \times 10^{-11}$ ergs cm$^{-2}$ s$^{-1}$ sr$^{-1}$
for the flat model, and
$4.5 \times 10^{-11}$ ergs cm$^{-2}$ s$^{-1}$ sr$^{-1}$
for the closed model.
We note that the numbers are only slightly smaller than the values
ignoring absorption (\S~\ref{sec:noabs} above).
Unlike the ionizing background from quasars, most of the diffuse emission
from clusters is produced at relatively small redshifts
(\S~\ref{sec:redshift} below), where the IGM opacity is low.

The predicted spectrum of the EUV background from clusters at the present
epoch $z_o = 0$ is shown in Figure~\ref{fig:spec_lum} for each of
our three cosmological models.
The ``bumps'' in these curves are due to 
absorption by \ion{H}{1} and \ion{He}{2}.
Clearly, absorption has only a small effect on the spectrum (compare
with Figure~\ref{fig:spec_lum_noabs}).
The present day ionization rates for cluster EUV models with IGM absorption
are
$\zeta_H = 1.6 \times 10^{-17}$,
$1.5 \times 10^{-17}$,
$8.2 \times 10^{-18}$ 
and
$\zeta_{HeII} = 1.6 \times 10^{-18}$, 
$1.4 \times 10^{-18}$,
$8.5 \times 10^{-19}$ s$^{-1}$ for the
open, flat, and closed cosmological models, respectively.  Ionization
rates from QSOs are given below for comparison (\S~\ref{sec:redshift}). 

Figure~\ref{fig:compare_lum} compares the predicted contributions of
clusters and quasars to the diffuse EUV background at $z_o = 0$ for the
open cosmological model.
The quasar results are taken from Haardt \& Madau (1996),
Figure 5a.
A comparison of the two curves shows that
the contribution to the EUV background from clusters is
$ < \frac{1}{2}\%$ of that of quasars.
Thus, while clusters do make a contribution to the EUV background at the
present epoch, the predominant source is due to AGN.

\subsubsection{Redshift Evolution} \label{sec:redshift}

The spectrum of the diffuse radiation field due to cluster IC EUV is plotted
as observed at redshifts of $z_o = 0$, 0.5, and 1.0
for the open cosmological model in Figure~\ref{fig:redshift}.
Models were calculated for higher redshifts ($z \ge 1.5$), but the
values much smaller than those shown.
This figure clearly shows that the intensity of the EUV background from
clusters decreases strongly with increasing redshift,
which means that the largest contribution to the
EUV background from clusters occurs at the current epoch.
On the other hand, the predicted intensity of the diffuse UV radiation field
due to quasars increases with increasing redshifts out to $z_o \sim 3$.
(Haardt \& Madau 1996, Figure 5).
Thus, the fractional contribution to the EUV background from clusters as
compared to quasars also decreases strongly with redshift.
This is because the abundance of rich clusters increases with time due
to gravitational mergers of smaller systems, while quasars were most
abundant in the early universe (at $z \ga 2$).

The ionization rate due to cluster EUV also decreases rapidly with
increasing redshift.
We find rates of
$\zeta_H = 1.6 \times 10^{-17}$,
$2.0 \times 10^{-18}$,
$1.2 \times 10^{-19}$,
$3.9 \times 10^{-21}$ s$^{-1}$
for \ion{H}{1} and
$\zeta_{HeII} = 1.6 \times 10^{-18}$,
$2.1 \times 10^{-19}$,
$1.3 \times 10^{-20}$,
$4.7 \times 10^{-22}$ s$^{-1}$
for \ion{He}{2}
at
$z_o = 0$, 0.5, 1.0, and 1.5, respectively,
for the
open cosmological model.
To compare these results with those from QSOs,
we use the results given by Haardt \& Madau (1996, Fig.~6) for an
open cosmological model with $q_0=0.1$, which give
$\zeta_H = 4.1 \times 10^{-14}$,
$1.6 \times 10^{-13}$,
$4.5 \times 10^{-13}$,
$9.3 \times 10^{-13}$ s$^{-1}$ for \ion{H}{1} 
and
$\zeta_{HeII} = 4.1 \times 10^{-16}$,
$1.6 \times 10^{-15}$,
$4.4 \times 10^{-15}$,
$9.0 \times 10^{-15}$ s$^{-1}$ for \ion{He}{2} 
at
$z_o = 0$, 0.5, 1.0, and 1.5, respectively.
In general, the ionization rates due to cluster EUV are $<$1\% of
those due to QSOs.
The largest contribution is for the \ion{He}{2} ionization rate at
$ z = 0$, where the cluster EUV gives $\sim$0.4\% of the quasar
rate.

\section{Conclusion} \label{sec:conclusion}

Observations with {\it EUVE} indicate that some clusters of galaxies are
luminous sources of extreme ultraviolet (EUV) radiation.
In order to limit the contribution of EUV emission from clusters to
the extragalactic background, we have assumed that this emission is
a common feature of clusters.
A promising theory for this emission is that it is due to
IC scattering of the Cosmic Microwave Background photons by relativistic
electrons. 
We have given a simple model for the average EUV luminosity expected from
clusters based on this theory.
We summed this emission over clusters of varying mass at different redshifts,
using the Press-Schechter formalism to determine the mass function of
clusters as a function of redshift.
We determined the amount of background radiation produced by clusters
through EUV IC emission.
The total mean intensity, spectrum, and the ionization rates for
\ion{H}{1} and \ion{He}{2} were determined at present and at a variety of
redshifts.
Because clusters form by the merger of smaller subclusters and the
abundance of massive clusters increases with time,
the amount of EUV background radiation should be larger at present than
in the past.
We compared our results to the ionizing background expected from quasars.
We find that while clusters do contribute a significant EUV background,
it is less than a percent of that expected from quasars.
Of course, this last conclusion is strengthened if EUV emission is not a
general feature of clusters.

\acknowledgements
We thank Avi Loeb and Ue-Li Pen for a useful conversation.
Support for this work was provided by the National Aeronautics and Space
Administration through Chandra Award Number GO0-1019X issued by
the Chandra X-ray Observatory Center, which is operated by the Smithsonian
Astrophysical Observatory for and on behalf of NASA under contract
NAS8-39073.

\clearpage

\clearpage

\clearpage

\begin{figure*}[p]
\vskip5.3truein
\includegraphics{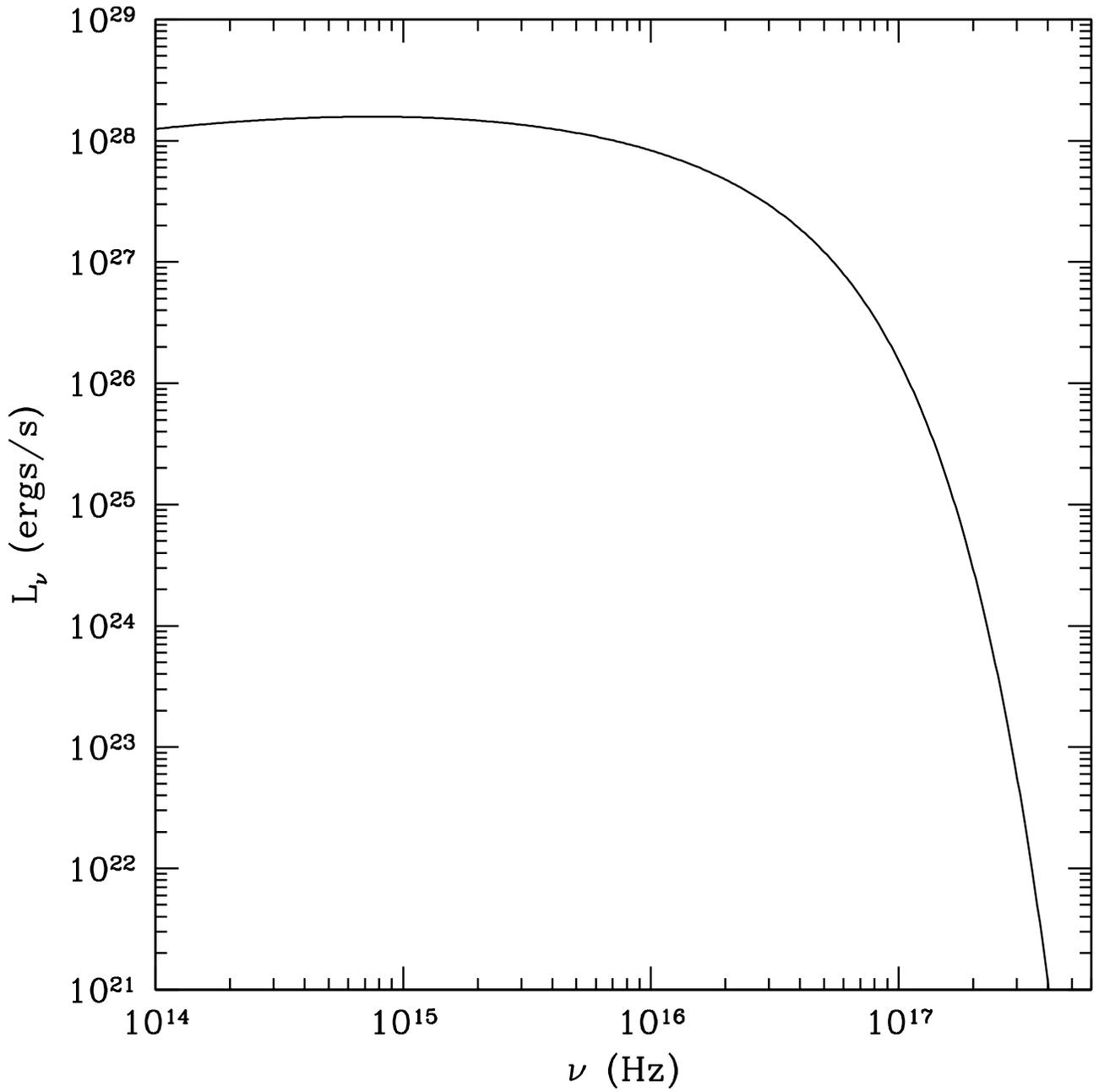}
\caption{The adopted IC EUV spectrum for our clusters, which
is the spectrum of Model~11 in Sarazin (1999).
Here, $L_\nu$ is the luminosity per unit frequency $\nu$.
This model has a total IC luminosity of $L_{IC} = 2.86 \times 10^{44}$
ergs s$^{-1}$.
The normalization of the spectrum of each model cluster is scaled to give
the correct value of $L_{IC}$ for that cluster.
\label{fig:spectrum_model}}
\end{figure*}

\clearpage

\begin{figure*}[p]
\vskip5.3truein
\includegraphics{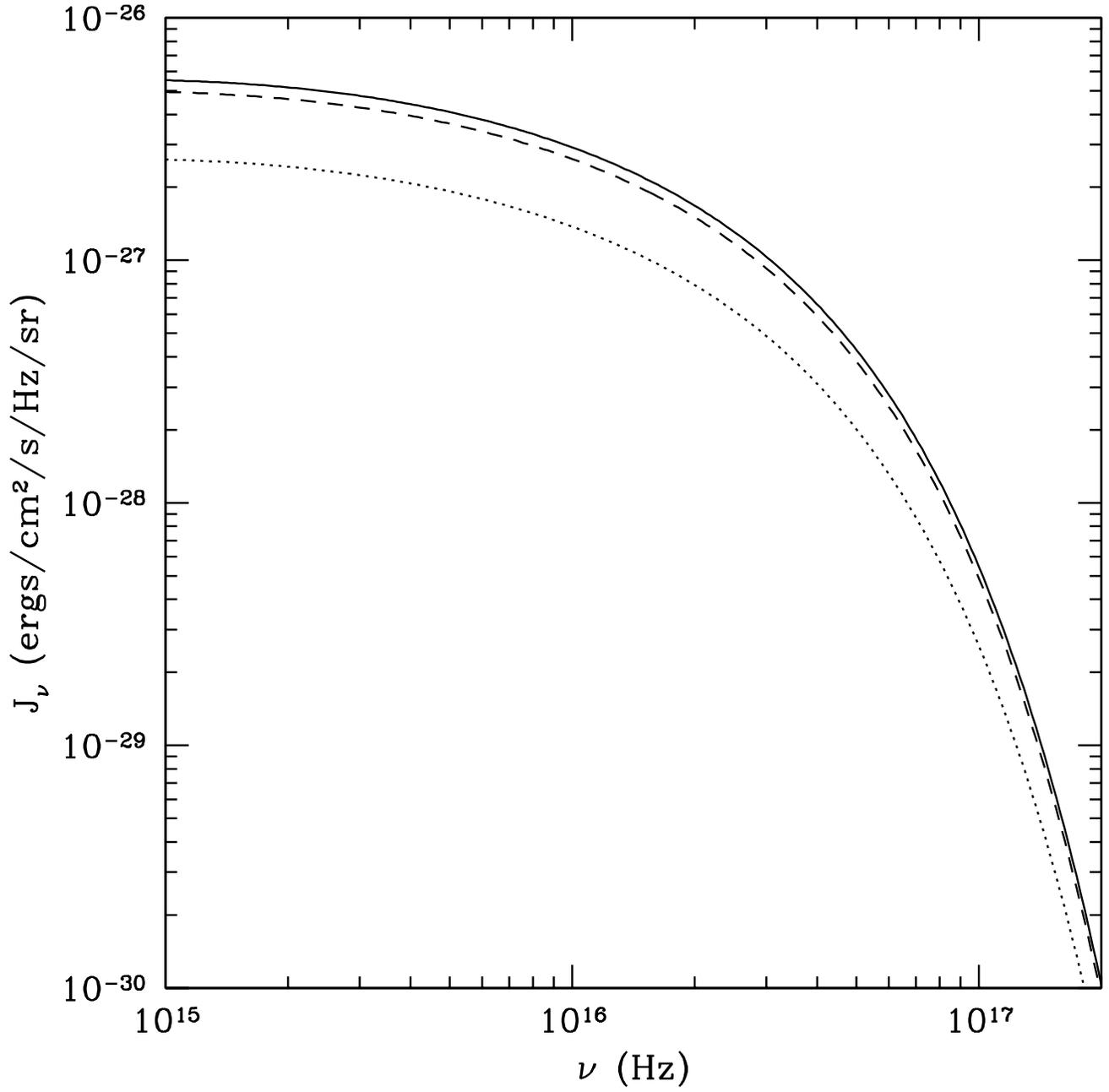}
\caption{
The mean specific intensity of the EUV background expected from
clusters at the current epoch for three distinct cosmologies.  The
solid curve represents an open Universe ($\Omega_0 = 0.3 \, , \,
\Omega_{\Lambda}=0 $), the dashed curve a flat Universe ($\Omega_0 =
0.3 \, , \,
\Omega_{\Lambda} = 0.7$), and the dotted curve a closed Universe
($\Omega_0 = 1.0 \, , \,
\Omega_{\Lambda} = 0$).  Absorption effects are ignored.
\label{fig:spec_lum_noabs}}
\end{figure*}

\clearpage

\begin{figure*}[p]
\vskip5.3truein
\includegraphics{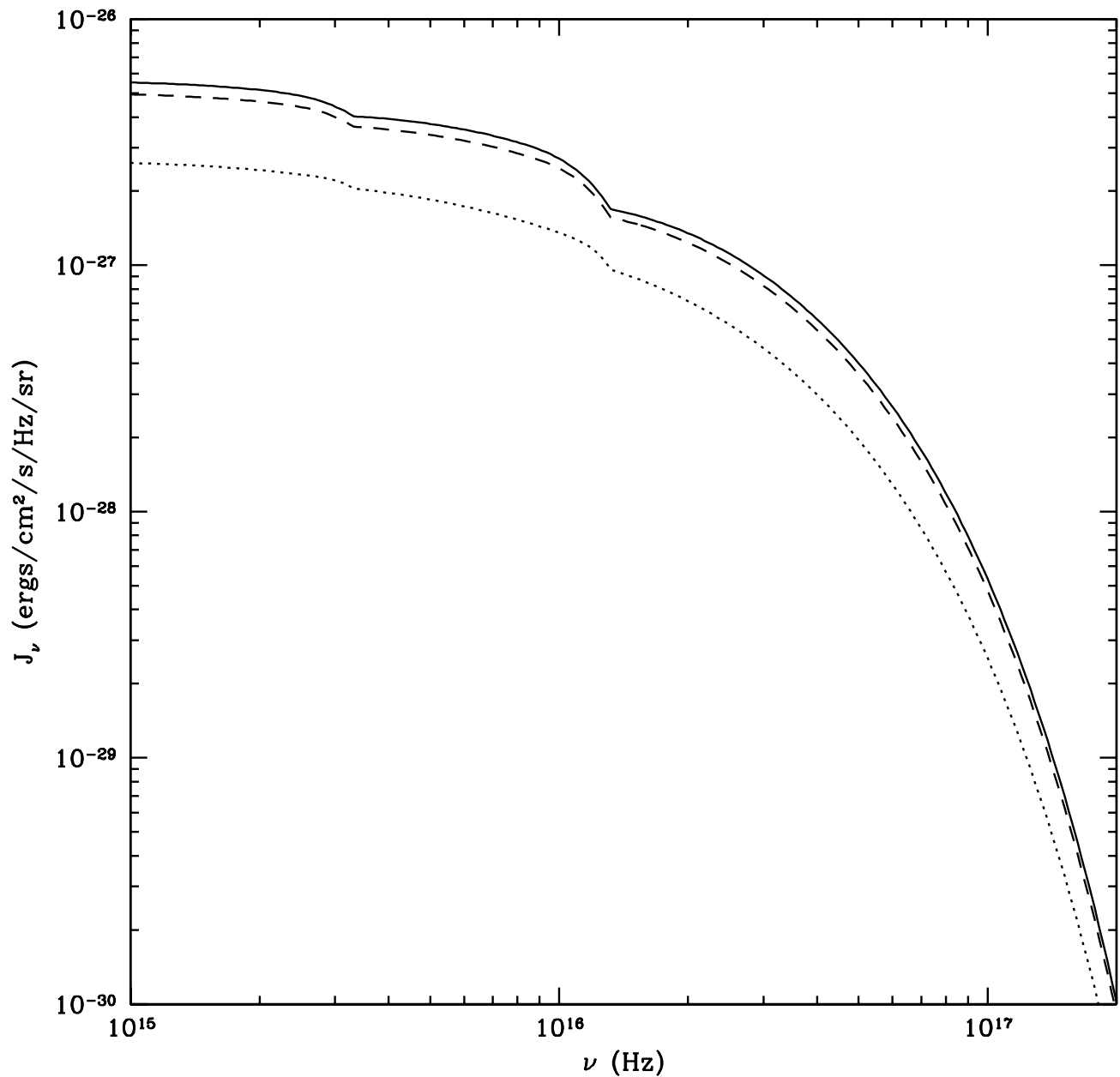}
\caption{
The mean specific intensity of the EUV background predicted
by models including IGM absorption by \ion{H}{1} and \ion{He}{2}.
The notation is the same as in Figure~\protect\ref{fig:spec_lum_noabs}.
\label{fig:spec_lum}}
\end{figure*}

\clearpage

\begin{figure*}[p]
\vskip5.3truein
\includegraphics{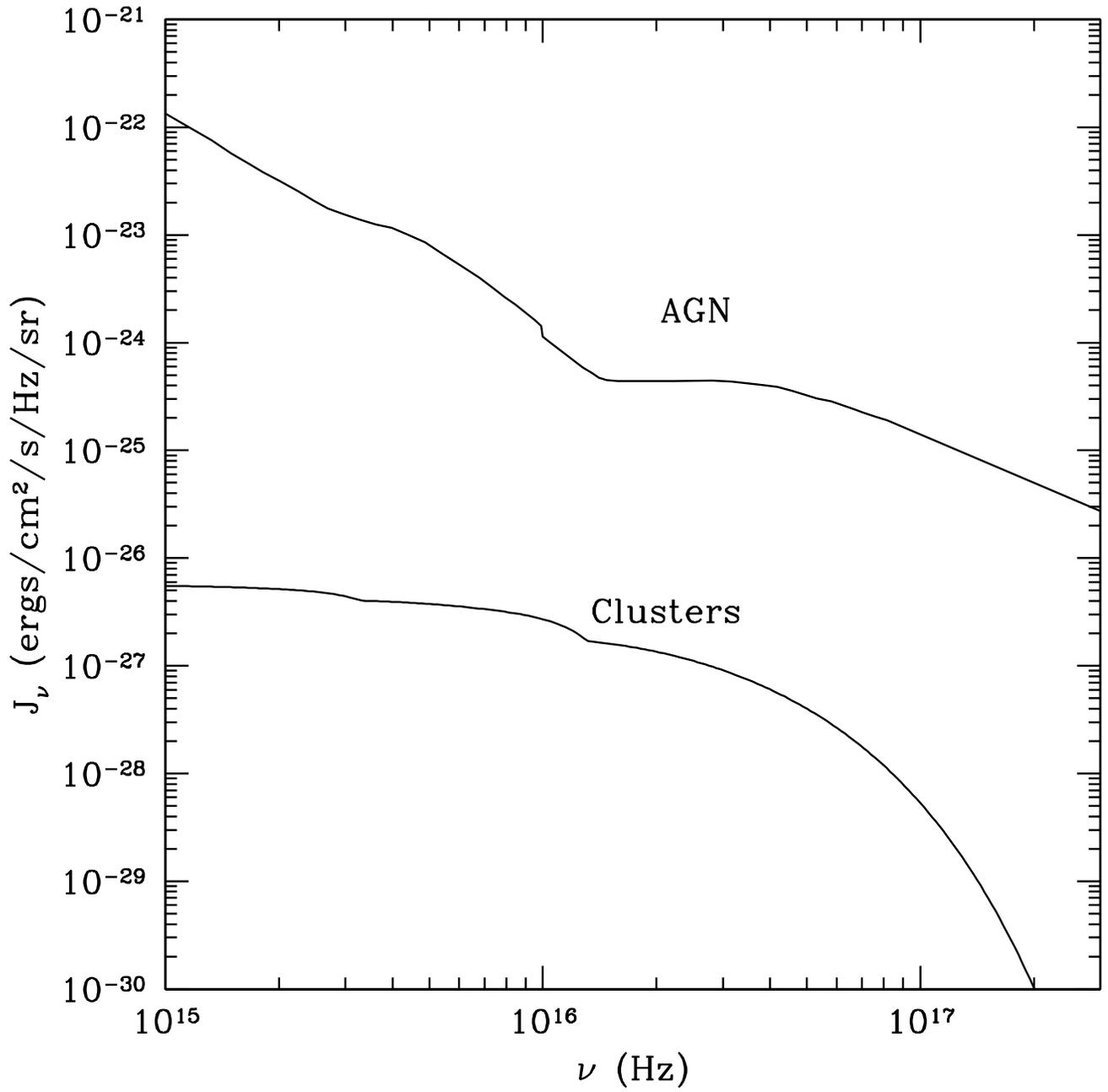}
\caption{
The predicted mean specific intensity of the EUV background at $z=0$
for an open Universe.
The curve labeled ``Clusters'' shows the contribution from
clusters while the curve labeled ``AGN'' shows the contribution from quasars.
\label{fig:compare_lum}}
\end{figure*}

\clearpage

\begin{figure*}[p]
\vskip5.3truein
\includegraphics{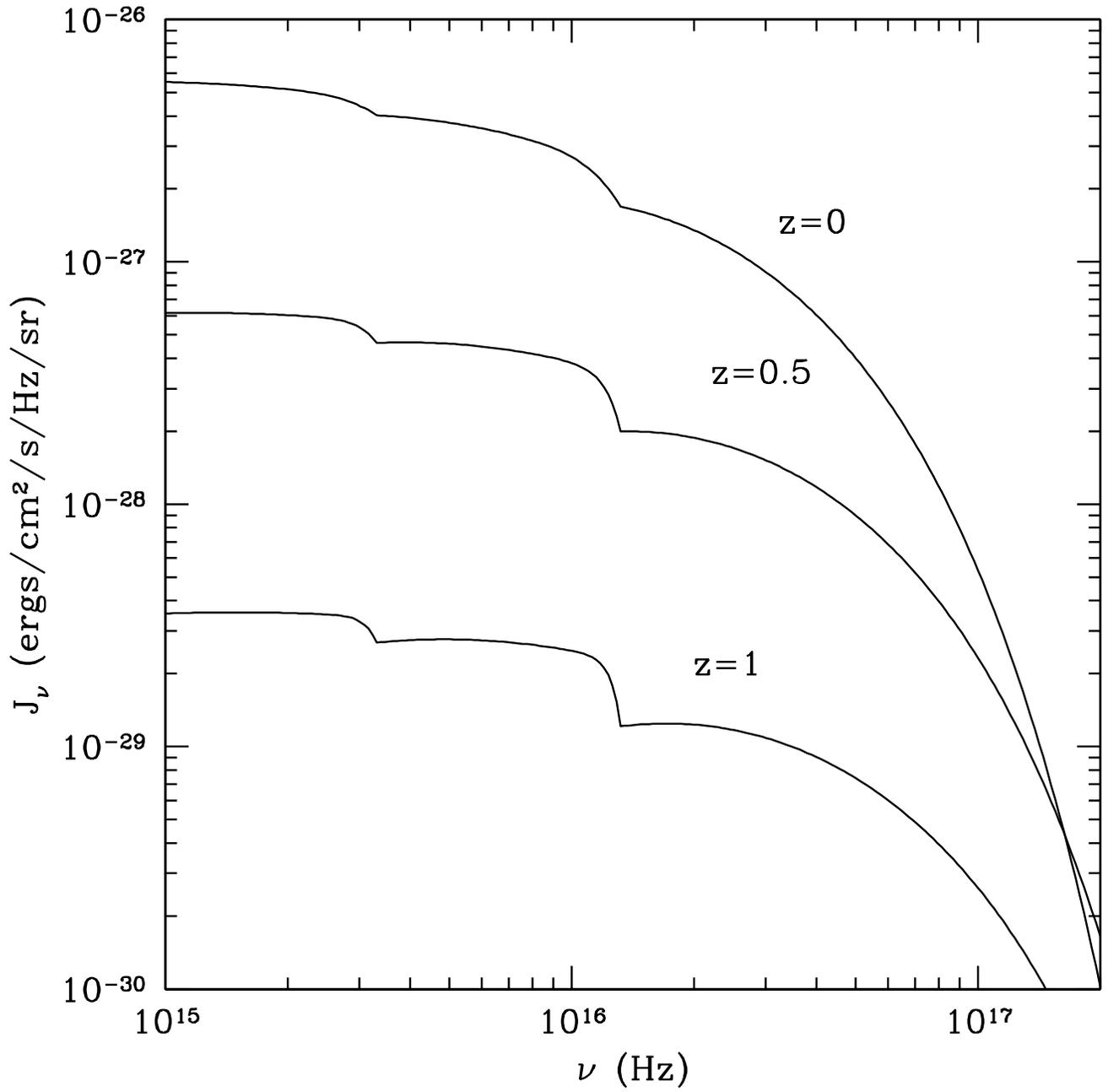}
\caption{
The spectrum of the diffuse EUV background from clusters as observed at
redshifts $z=0$, $z=0.5$, and $z=1$ for the open cosmological model
including IGM absorption.
\label{fig:redshift}}
\end{figure*}

\end{document}